\input{aipcheck.tex}
\documentclass[final]{aipproc}

\layoutstyle{6x9}

\newcommand\doingARLO[2][]{%
  \ifx\mmref\undefined #1\else #2\fi
}

\def\Mpl{M_\mathrm{P}}

\newcommand{\gtrsim}{ \mathop{}_{\textstyle \sim}^{\textstyle >} }
\newcommand{\lesssim}{ \mathop{}_{\textstyle \sim}^{\textstyle <} }

\begin{document}

\title
      {Gravitino Dark Matter}

\classification{98.70.Sa, 95.35.+d}
\keywords{supergravity, dark matter, leptogenesis, cosmic rays}

\author{Wilfried Buchm\"uller}{
  address={Deutsches Elektronen-Synchrotron DESY, 22607 Hamburg, Germany},
  email={buchmuwi@mail.desy.de},
}

\iftrue

\fi

\copyrightyear  {2001}

\begin{abstract}
\noindent
Gravitino dark matter, together with thermal leptogenesis, implies an upper
bound on the masses of superparticles. In the case of broken R-parity the 
constraints from primordial nucleosynthesis are naturally satisfied and
decaying gravitinos lead to characteristic signatures in high energy cosmic
rays. Electron and positron fluxes from gravitino decays cannot explain both,
the PAMELA positron fraction and the electron $+$ positron flux recently
measured by Fermi LAT. The observed fluxes require astrophysical sources. 
The measured antiproton flux allows for a sizable contribution of decaying 
gravitinos to the gamma-ray spectrum, in particular a line at an energy 
below 300~GeV. 
\end{abstract}

\date{\today}

\maketitle

\section{WHY GRAVITINO DARK MATTER?}

An unequivocal prediction of locally supersymmetric extensions of the Standard 
Model is the gravitino, the gauge fermion of supergravity \cite{fnf76}. Its 
discovery would be of fundamental importance, as the discovery of the W- and
Z-bosons for the electroweak theory. Depending on the mechanism of 
supersymmetry breaking, it can be the lightest superparticle (LSP), which makes
it a natural dark matter candidate. So far, almost nothing is known about the 
gravitino mass. Examples of considered mass ranges are
\begin{itemize}
\item
$m_{3/2} < 1\ \mathrm{keV}$, corresponding to hot dark matter \cite{pp81},
\item
$1\ \mathrm{keV} < m_{3/2} < 15\ \mathrm{keV}$, 
representing warm dark matter \cite{gkr08},
\item
$100\ \mathrm{keV} < m_{3/2} < 10\ \mathrm{MeV}$, 
a range of cold dark matter, favoured by gauge mediation and thermal 
leptogenesis \cite{fiy03,htv08},
\item
$100\ \mathrm{GeV} < m_{3/2} < 1\ \mathrm{TeV}$, a range of cold dark 
matter typical for gaugino and gravity mediation, suggested in connection 
with thermal leptogenesis \cite{bbp98}.
\end{itemize}

It is very interesting that the various gravitino mass ranges are strongly 
correlated with the masses of other superparticles and with the history of 
the cosmological evolution. This leads to upper bounds on the gluino mass,
lower and upper bounds on the next-to-lightest superparticle mass (NLSP) 
and also upper
bounds on the reheating temperature in the early universe. Moreover, there
is a close connection with baryogenesis.

It would be very exciting to discover a massive gravitino and to establish in
this way spontaneously broken local supersymmetry as a fundamental, hidden 
symmetry of nature. This appears impossible for the most popular scenario of
heavy gravitinos with neutralino LSP. On the contrary, it may be possible
to discover a gravitino LSP which manifests itself as missing energy in NLSP 
decays \cite{gr98,bhx04}.

\section{GRAVITINO PROBLEM OR VIRTUE?}

In a supersymmetric plasma at high temperature gravitinos are thermally
produced, mostly by QCD processes. Their number density $n_{3/2}$ increases 
linearly with the reheating temperature,
\begin{equation}
{n_{3/2}\over n_\gamma} \propto {\alpha_3\over M_{\rm p}^2}\,T_R\;,
\end{equation}
where $M_{\rm p}$ and $\alpha_3$ are the Planck mass and the QCD fine
structure constant, respectively. The late decay of heavy gravitinos 
alters the successful BBN prediction, which implies upper bounds for the
reheating temperature $T_R$. The most stringent one from hadronic decays
reads \cite{kkm05} 
\begin{equation}
T_R < {\cal O}(1)\times 10^5\ {\rm GeV}\ .
\end{equation}
This is clearly incompatible with thermal leptogenesis \cite{fy86} which leads 
to the lower bound $T_R \gtrsim 10^9~\mathrm{GeV}$ (cf.~\cite{di02,bdp04}).

The conflict between the upper bound from BBN and the lower bound from 
leptogenesis on the reheating temperature can be avoided if the gravitino
is the LSP \cite{bbp98}. In this case the BBN bounds apply to the NLSP which
is quasi-stable. One also has to worry about gravitino dark matter
which may overclose the universe.

In general, two processes contribute to the generation of gravitino
dark matter. In the SuperWIMP mechanism \cite{ckr99,frt03} gravitinos are 
produced in WIMP decays. The gravitino mass density is then determined by the 
NLSP density,
\begin{equation}
\Omega_{3/2} = {m_{3/2}\over m_{\rm NLSP}} \Omega_{\rm NLSP}\ ,
\end{equation}
which is independent of the reheating temperature $T_R$. The BBN constraints
require, however, rather large NLSP masses \cite{fst04},  which makes it 
difficult to test this mechanism at the LHC.
The thermal production of gravitinos is dominated
by $2 \rightarrow 2$ QCD scattering processes. Solving the Boltzmann equations
one obtains \cite{bbb00,ps06}
\begin{equation}
\Omega_{3/2}h^2\simeq 0.5 
\left(\frac{T_R}{10^{10}\ \rm{GeV}}\right)
\left(\frac{\rm{100\ GeV}}{m_{3/2}}\right)
\left(\frac{m_{\tilde{g}}(\mu)}{\rm{1\ TeV}}\right)^2\ ,
\end{equation}
where the two-loop running of the gluino mass from the rehating temperature
to the electroweak scale has been taken into account \cite{bes08}.
It is remarkable that the observed dark matter density is obtained for
typical SUSY breaking parameters, $m_{3/2} \sim 100\ {\rm GeV}$,
$m_{\tilde{g}} \sim 1\ {\rm TeV}$ and a reheating temperature characteristic
for leptogenesis, $T_R \sim \sqrt{m_{3/2}M_P} \sim 10^{10}\ {\rm GeV}$
\cite{bbp98}.

\begin{figure}[t]
\includegraphics[height=.3\textheight]{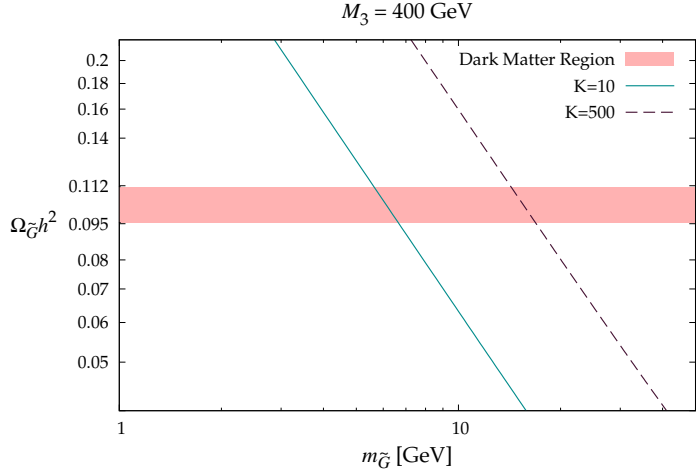}
\caption{Relic gravitino density as function of the gravitino mass; the 
GUT scale parameter $M_3 = 400\ \mathrm{GeV}$ corresponds to the
on-shell gluino mass $m_{\tilde{g}} \simeq 1\ \mathrm{TeV}$. The leptogenesis
parameters are $M_1 = 10^9\ \mathrm{GeV}$ and $K$ which lies
in the `strong washout regime'. From \cite{erbe}.}
\label{fig:erbe}
\end{figure}

The type and masses of NLSP's consistent with leptogenesis and gravitino dark 
matter are strongly restricted by constraints from BBN (cf.~\cite{ste08}), in 
particular by the catalyzed production of $^6$Li in the case of a stau NLSP
\cite{pos07}. If the reheating process, which leads to the temperature $T_R$,
is taken into account the constraints are sometimes relaxed. For instance,
in the case of inflaton decays into right-handed neutrinos, $\phi \rightarrow
N_1N_1 \rightarrow (lH)(lH) \rightarrow \ldots$, the reheating temperature
required for leptogenesis is smaller by about one order of magnitude
\cite{ahx99,hwp08}. If the Boltzmann equation for lepton asymmetry and 
gravitino density are solved simultaneously, a connection between gravitino
mass and neutrino parameters is obtained directly, without any reference
to the reheating temperature. As an example (see Figure~\ref{fig:erbe}), for 
$M_1 = 10^9\ \mathrm{GeV}$, $K=10$ ($\tilde{m}_1 = 0.01\ \mathrm{eV}$) and 
$m_{\tilde{g}} = 1\ \mathrm{TeV}$ one finds $m_{3/2} \simeq 6\ \mathrm{GeV}$
\cite{erbe}.

\section{DECAYING GRAVITINO DARK MATTER}

Nucleosynthesis, leptogenesis and gravitino dark matter can all be
consistent in the case of a small R-parity breaking \cite{bcx07}, which leads 
to the  processes $\widetilde\tau_R \rightarrow \tau\nu_\mu, \mu\nu_\tau$, 
{${\widetilde\tau_L}\rightarrow b^c t$}, ...
and also to $\psi_{3/2} \rightarrow \gamma \nu$. Small R-parity breaking 
couplings can be induced by B-L breaking,
\begin{eqnarray}
\lambda \sim  h^{(e,d)}\Theta \lesssim 10^{-7}\;, \quad
\Theta \sim {v_{B-L}^2\over \Mpl^2}\ ,
\end{eqnarray}
where $h^{(e,d)}$ are the lepton and down-quark Yukawa couplings, respectively.
The `short' NLSP lifetimes, e.g.,
\begin{eqnarray}
c \tau^{\mathrm{lep}}_{\tilde\tau} \sim
30~{\rm cm}\left(  \frac{m_{\tilde{\tau}}}{200 {\rm GeV}} \right)^{-1}
\left(\frac{\lambda}{10^{-7}}\right)^{-2}\ ,
\end{eqnarray}
typically lead to NLSP decay before BBN. One finds that BBN, thermal 
leptogenesis
and gravitino dark matter are consistent for $10^{-14} < \lambda,\lambda' < 
10^{-7}$ and { $m_{3/2} \gtrsim 5~{\rm GeV}$ \cite{bcx07}. 
Characteristic signals at the LHC can be strongly ionising macroscopic
charged tracks, followed by a muon track, or a jet and missing 
energy, corresponding to $\widetilde \tau \rightarrow \mu \nu_\tau$ and
$\widetilde\tau \rightarrow \tau\nu_\mu$, respectively.

\begin{figure}[t]
\hskip -1cm
\includegraphics[height=.3\textheight]{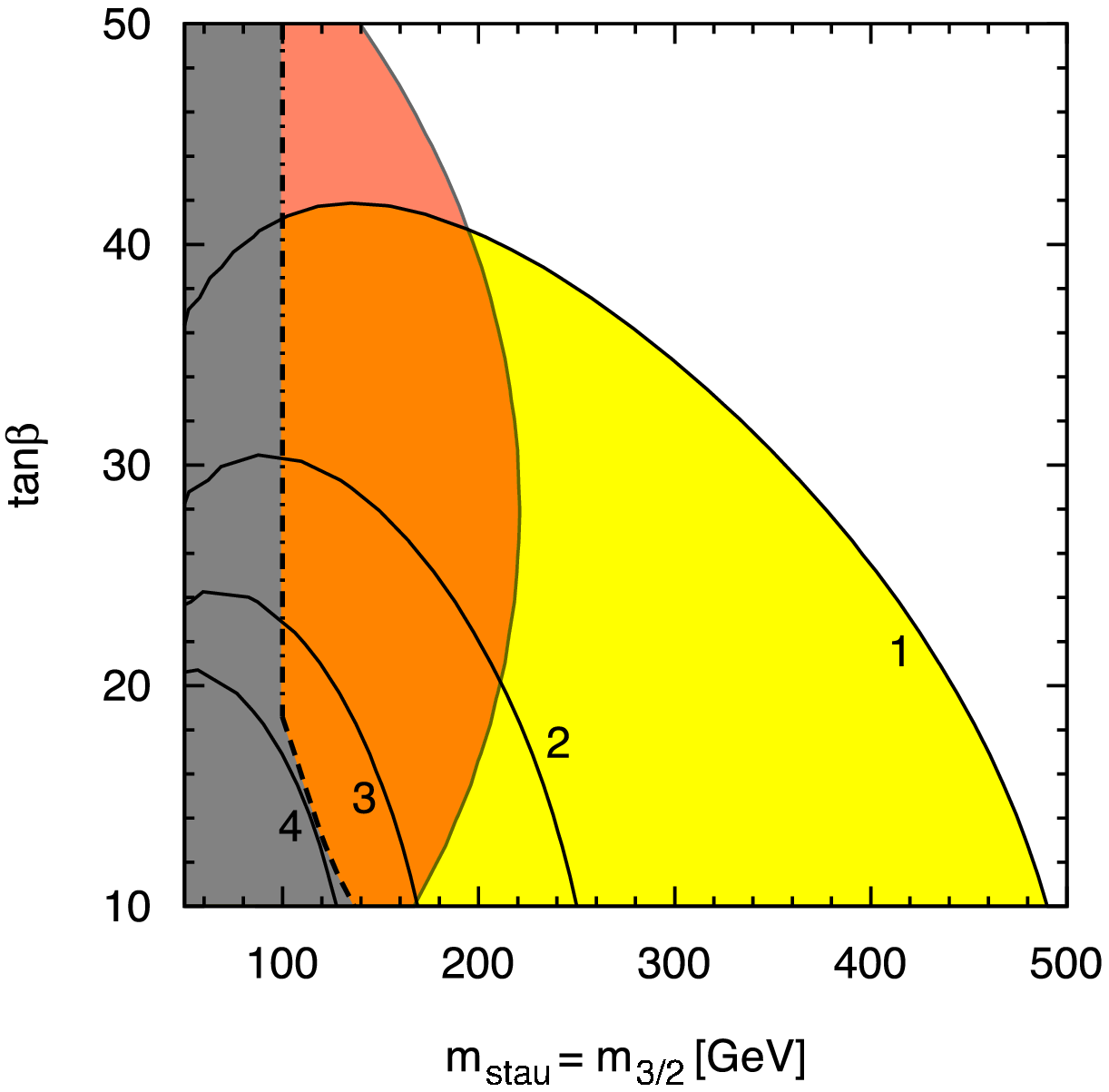}\hskip 1cm
\includegraphics[height=.3\textheight]{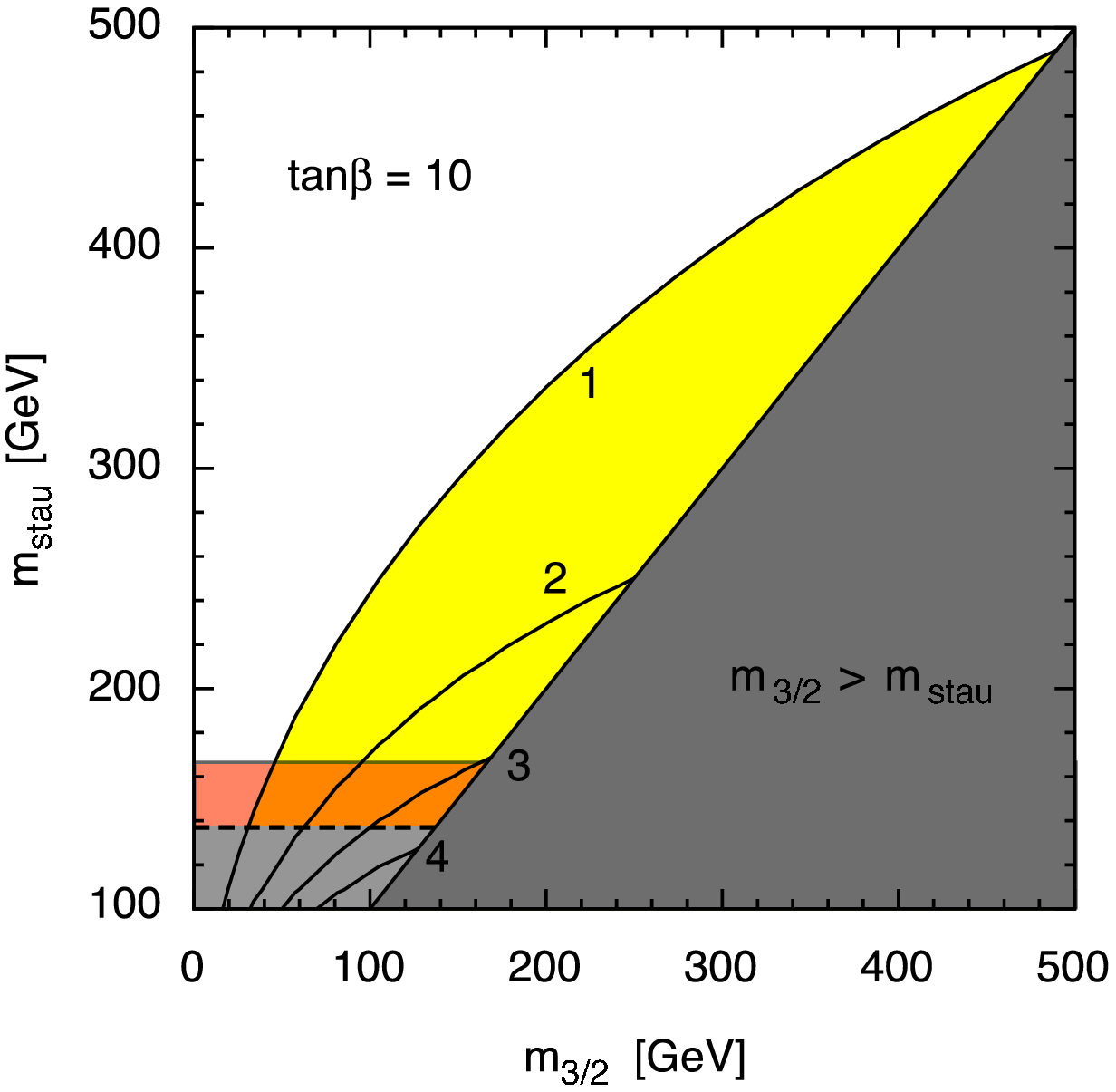}
\hskip -9cm {\small (a)}\hskip 7cm{\small (b)}
\caption{Boundary conditions (B) with stau NLSP.
(a) Contours of constant reheating temperature, $T_R=(1-4) \times 10^9$~GeV,
with $\Omega_{3/2} = \Omega_\mathrm{DM}$ (solid lines); the gray region is 
excluded by constraints from low-energy experiments;
thermal leptogenesis is possible in the yellow and orange regions; the
orange region is favored by the muon $g-2$ anomaly at the 2$\sigma$ level.
(b) Contours of constant reheating temperature in the 
$m_\mathrm{stau}-m_{3/2}$ plane; in the dark gray region, the gravitino is 
not the LSP. From \cite{bes08}.
}
\label{fig:bounds}
\end{figure}

The gravitino decay $\psi_{3/2} \rightarrow \gamma \nu$ is suppressed 
both by the Planck mass and small R-parity breaking couplings, so that
the lifetime is much longer than the age of the universe \cite{ty00},
\begin{eqnarray}
\tau_{3/2}\ \sim \ 10^{26}\ {\rm s}\  
\left(\frac{\lambda}{10^{-7}}\right)^{-2}
\left(\frac{m_{3/2}}{10~{\rm GeV}}\right)^{-3}\ .
\end{eqnarray}
Decaying dark matter with lifetime ${\cal O}(10^{26})\ {\rm s}$ has become very
popular after the recent observation of the rise in the cosmic-ray positron 
fraction by the PAMELA collaboration \cite{pamela08}. Note that R-parity 
breaking scenarios with gravitino lifetimes of this order can be realized
in various ways \cite{lor07,jmx08,es09,clx09,fs09}.  

The consistency of leptogenesis and gravitino dark matter implies 
important constraints on the superparticle mass spectrum, which depends on
the boundary conditions of the supersymmetry breaking parameters at the 
grand unification (GUT) scale. Typical examples are
\begin{eqnarray}
  (\mathrm{A})~~~m_0 = m_{1/2},~~a_0 = 0\ ; \qquad
  (\mathrm{B})~~~m_0 = 0,~~m_{1/2},~~a_0 = 0\ ,
\end{eqnarray}
with a bino-like neutralino (A), and  
a right-handed stau (B) as NLSP, respectively. The corresponding upper
bounds on the gravitino and stau masses are shown in 
Figure~\ref{fig:bounds} for reheating temperatures in the range
$T_R = (1-4)\times 10^9\ \mathrm{GeV}$ \cite{bes08}. Relaxing the boundary 
conditions at
the GUT scale, gravitino masses up to $1.4\ \mathrm{TeV}$ are possible 
\cite{hty09}.

\section{CONSTRAINTS FROM PAMELA AND FERMI}

The R-parity violating gravitino decays $\psi_{3/2} \rightarrow \gamma\nu, 
h\nu, Z\nu,W^\pm l^\mp$ lead to interesting cosmic-ray signatures \cite{tran}
which we shall discuss in the following in a model-independent way
based on an operator analysis \cite{bix09}. 
The mass scales multiplying the non-renormalizable operators 
are inverse powers of the Planck mass $\Mpl$ and the supersymmetry breaking 
gravitino mass $m_{3/2}$. This assumes for the masses $m_{\mathrm{SM}}$ of 
Standard Model particles, the gravitino mass and the masses 
$m_{\mathrm{soft}}$ of other superparticles the hierarchy 
$m^2_{\mathrm{SM}} \ll m^2_{3/2} \ll m^2_{\mathrm{soft}}$. One easily verifies 
that the dimension-5 and dimension-6 operators for the $R$-parity violating
couplings of the gravitino are given by
\begin{eqnarray}
{\cal L}_{\mathrm{eff}} = 
\frac{i\kappa}{\sqrt{2}M_{\rm P}}
\left\{\bar{l}\gamma^\lambda\gamma^\nu D_\nu \phi \psi_{\lambda} 
+\frac{i}{2}\bar{l}\gamma^\lambda
\left(\xi_1 g'YB_{\mu\nu} + \xi_2 g W_{\mu\nu}\right)
\sigma^{\mu\nu}\phi\psi_{\lambda} \right\} + \mathrm{h.c.}\ , 
\end{eqnarray}
where typically $\xi_{1.2} = {\cal O}(1/m_{3/2})$. Note, that
in general $\kappa$ and the product $\kappa \xi_{1,2}$ are independent 
parameters which depend on flavour. 

\begin{figure}[t]
\includegraphics[height=.3\textheight]{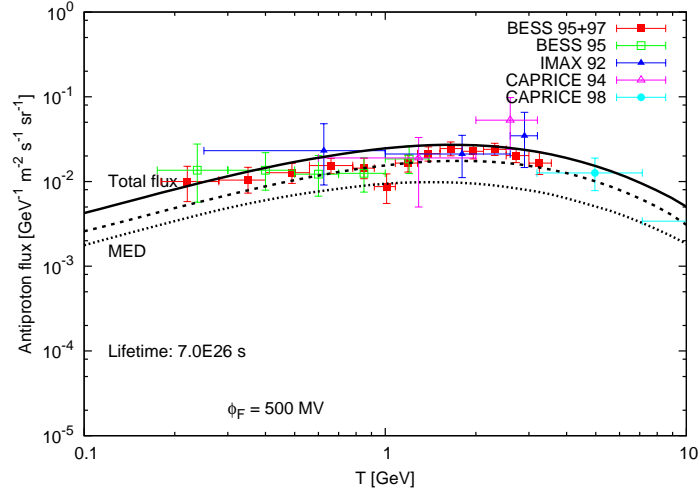}
\caption{
Antiproton flux for $m_{3/2} = 200~\mathrm{GeV}$ in the 
MED set of propagation parameters that saturate the antiproton 
overproduction bound. Dotted line: antiproton 
flux from gravitino decays, dashed line: secondary antiproton flux from 
spallation in the case of minimal nuclear cross section, solid line: 
total antiproton flux. The gravitino lifetime is 
$\tau_{3/2} = 7 \times 10^{26}~\mathrm{s}$. From \cite{bix09}.
}
\label{fig:pbar}
\end{figure}

The dimension-5 operator describes the processes 
$\psi_{3/2} \rightarrow h\nu, Z\nu, W^\pm l^\mp$ which,
after fragmentation of Higgs, Z- and W-bosons, yields continuous
gamma and antimatter spectra, $\psi_{3/2} \rightarrow \gamma X, \bar{p} X, 
e^+ X$, which therefore are strongly correlated. It is remarkable that
the decay $\psi_{3/2}\rightarrow \gamma\nu$ is controlled by the 
dimension-6 operator. Hence, the intensity of a line in the gamma-ray spectrum
is not tied to the continuous part of the spectrum. 

The interstellar antiproton flux from gravitino decay suffers from 
uncertaintes in the determination of the physical parameters in 
the propagation of charged cosmic rays in the diffusive halo, 
leading to uncertainties in the magnitude of fluxes as large
as two orders of magnitude at the energies relevant
for present antiproton experiments. The requirement that the total 
antiproton flux from gravitino decay be consistent with 
measurements gives a lower bound on the gravitino mass
which strongly depends on the choice of the halo model.
In the following we shall adopt the MED propagation model,
which provides the best fit to the B/C ratio and measurements of 
flux ratios of radioactive cosmic-ray species~\cite{Maurin:2001sj}.

A conservative upper bound on the antiproton flux from gravitinos is obtained
by demanding that the total flux is not larger than the theoretical uncertainty
band of the MED propagation model. This means that a `minimal' dark matter
lifetime for the MED model can be defined by a scenario where the 
secondary antiproton flux from spallation is 25\% smaller
than the central value, due to a putative overestimation of
the nuclear cross sections, and the total antiproton flux saturates
the upper limit of the uncertainty band which stems from astrophysical 
uncertainties discussed above. 

Using this prescription one finds the lower bound on the gravitino lifetime  
$\tau^\mathrm{min}_{3/2} \simeq 7\times 10^{26}~\mathrm{s}$ for 
$m_{3/2} = 200~\mathrm{GeV}$.
The corresponding antiproton flux from gravitino decay, the
secondary antiproton flux from spallation and the total antiproton flux
are shown in Figure~\ref{fig:pbar} together with the experimental measurements
by BESS, IMAX and WiZard/CAPRICE, and the uncertainty band 
from the nuclear cross sections in the MED propagation model.
The minimal lifetime $\tau^\mathrm{min}_{3/2}$ can be compared with the 
gravitino lifetime needed to explain the PAMELA positron fraction excess.

It is now straightforward to calculate the positron flux at Earth from
gravitino decay in the MED propagation model~\cite{Maurin:2001sj}.
Note that the sensitivity of the positron fraction 
to the propagation model is fairly mild at the energies where
the excess is observed, since these positrons are produced
within a few kiloparsecs from the Earth and barely suffer
the effects of diffusion.

To compare the predictions
to the PAMELA results, we calculate the positron fraction,
defined as the flux ratio $\Phi_{e^+}/(\Phi_{e^+}+\Phi_{e^-})$. 
For the background fluxes of primary and secondary electrons, 
as well as secondary positrons, we extract the fluxes from 
``Model 0'' presented by the Fermi LAT
collaboration in \cite{gxx09}, which fits well the low energy data points of
the total electron plus positron flux and the positron fraction,
and is similar to the MED model for energies
above a few GeV \cite{Delahaye:2008ua}.
Then, the positron fraction reads
\begin{equation}
{\rm PF}(T) = \frac{\Phi_{e^+}^{\rm{DM}}(T) + \Phi_{e^+}^{\rm{bkg}}(T)}
{\Phi_{e^+}^{\rm{DM}}(T) + \Phi_{e^+}^{\rm{bkg}}(T) +\Phi_{e^-}^{\rm{DM}}(T)
+ k \;  \Phi_{e^-}^{\rm{bkg}}(T) },
\label{PF}
\end{equation}
where $k = {\cal O}(1)$ is the normalization of the 
astrophysical contribution to the primary electron flux,
which is chosen to provide a qualitatively good fit to the data.

For the PAMELA positron excess to be due to gravitino dark matter decay, the 
gravitino mass must be at least 200 GeV. The decay $\psi_{3/2} \rightarrow 
W^\pm \ell^\mp$ then has a branching ratio of $\sim$ 50\%, and the hard leptons
that are directly produced in these decays may account for the rise in the 
positron fraction if a significant fraction of these leptons has 
electron or muon flavour. 

Consider first the extreme case that the decays occur purely into electron 
flavour. For $m_{3/2}=200~\mathrm{GeV}$, the PAMELA excess can then be 
explained for the gravitino lifetime 
$\tau^e_{3/2}(200) \simeq 3.2\times 10^{26}~\mathrm{s}$,
as illustrated by Figure~\ref{fig:pf1}. 
Note that this lifetime is a factor 2 smaller than the minimum lifetime 
which was obtained from the antiproton constraint. 
However, the MIN model and other sets of parameters that 
yield intermediate values for the antiproton flux can easily be 
compatible with both the positron fraction and the antiproton-to-proton 
ratio observed by PAMELA. The situation is very similar for 
$m_{3/2}=400~\mathrm{and}~600~\mathrm{GeV}$.

\begin{figure}[t]
\hskip -1.5cm
\includegraphics[height=.25\textheight]{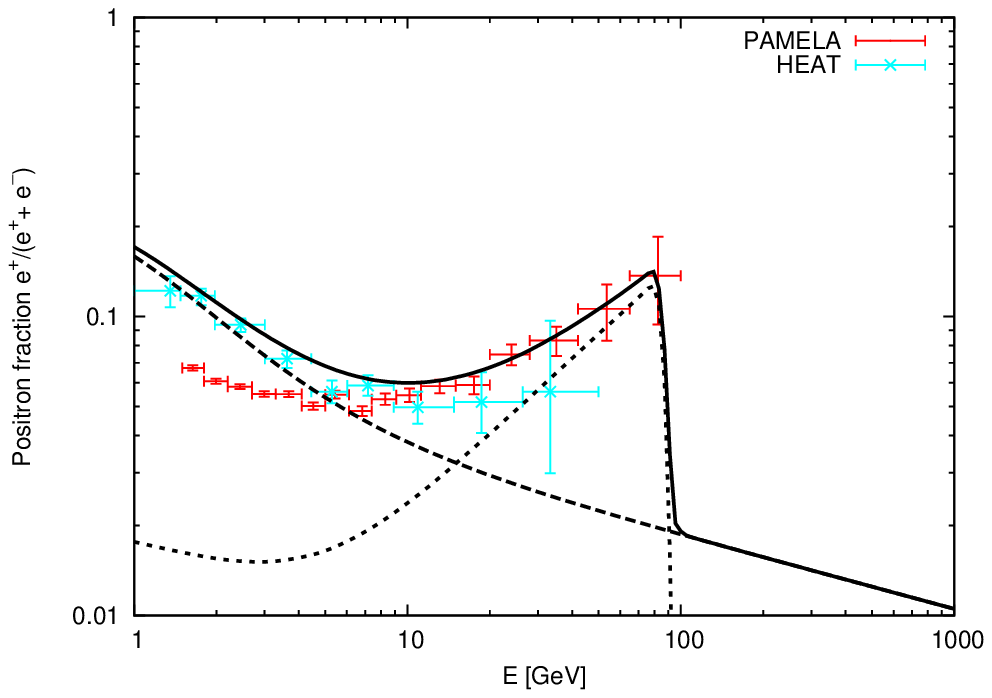}\hskip 1cm
\includegraphics[height=.25\textheight]{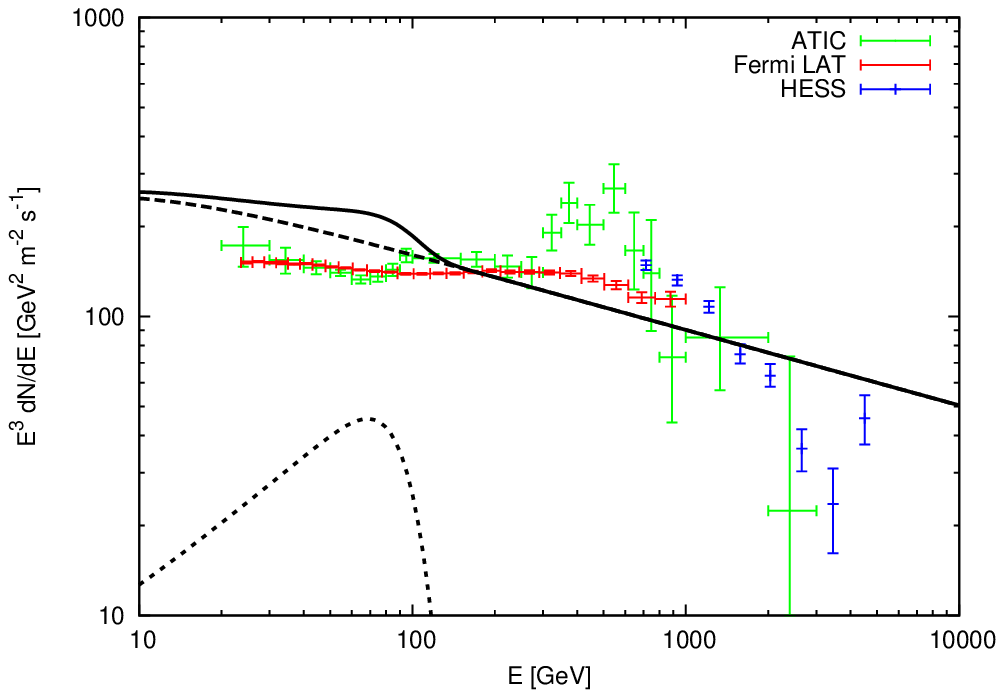}
\hskip -10.5cm {\small (a)}\hskip 8.5cm{\small (b)}
\caption{`Electron dominance'.
Contribution from dark matter decay to the positron fraction (a), and the total
electron $+$ positron flux (b), compared with data from PAMELA and HEAT, and
ATIC, Fermi LAT and HESS, respectively; 
$m_{3/2} = 200~\mathrm{GeV},~\tau_{3/2} = 3.2 \times 10^{26}~\mathrm{s}$; for
$W^{\pm}l^{\mp}$ decays pure electron flavour is assumed. 
The ``Model 0'' \cite{gxx09} background is used, and for comparison with 
Fermi LAT data 25\% energy resolution is taken into account. From \cite{bix09}.
}
\label{fig:pf1}
\end{figure}
\begin{figure}[b]
\hskip -1.5cm
\includegraphics[height=.25\textheight]{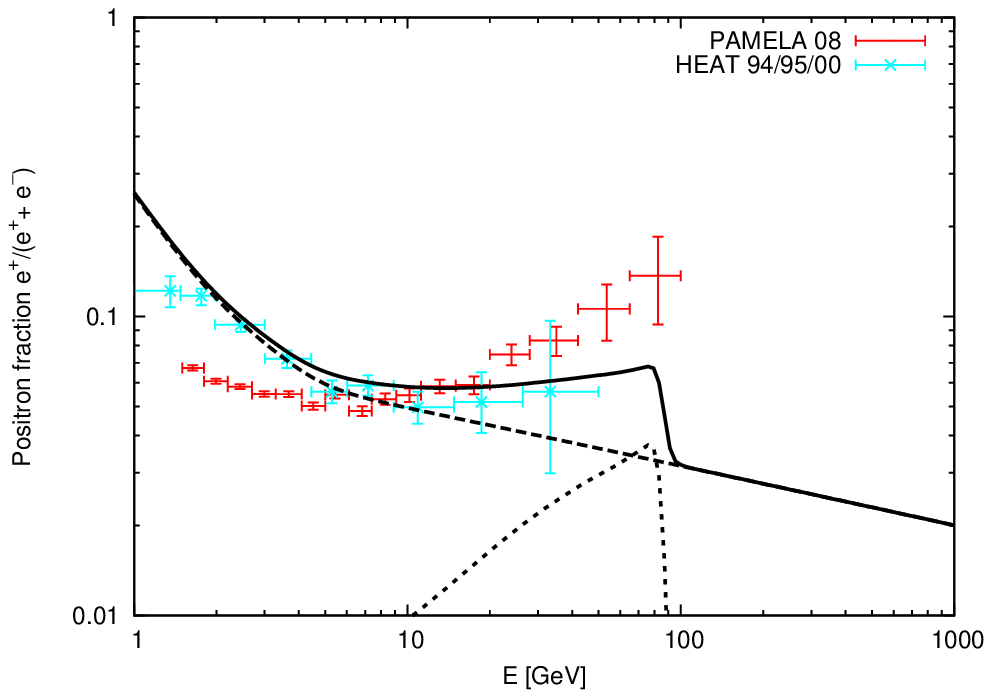}
\hskip 1cm
\includegraphics[height=.25\textheight]{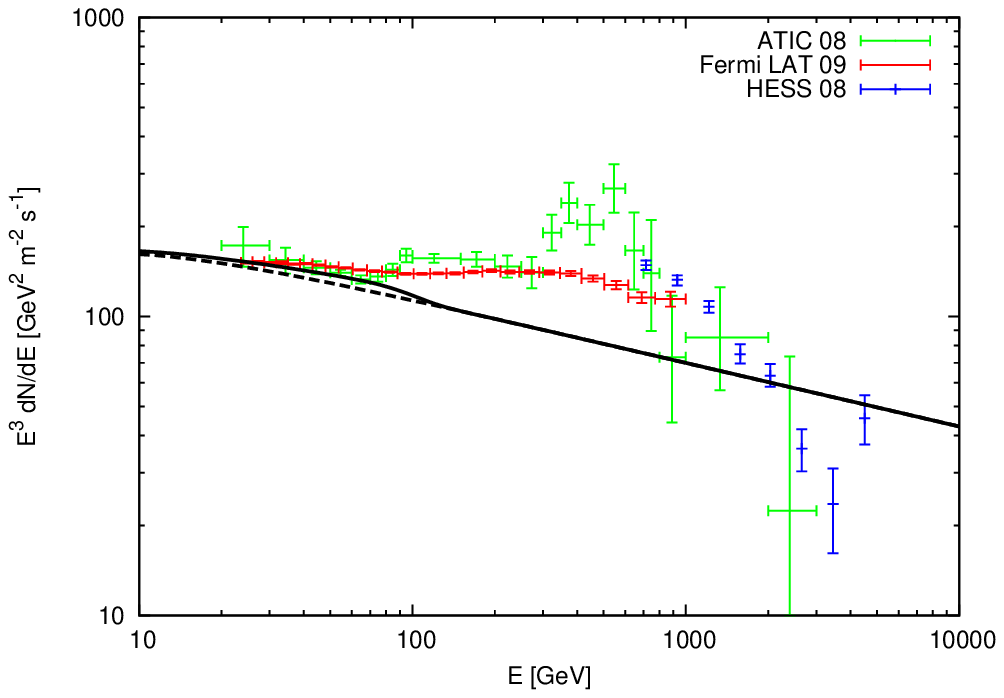}
\hskip -10.5cm {\small (a)}\hskip 8.5cm{\small (b)}
\caption{`Flavour democracy'.
Contribution from dark matter decay to the positron fraction (a), and the total
electron $+$ positron flux (b), compared with data from PAMELA and HEAT, and
ATIC, Fermi LAT and HESS, respectively; 
$m_{3/2} = 200~\mathrm{GeV}$, with the minimal lifetime 
$\tau_{3/2} = 7 \times 10^{26}\ \mathrm{s}$; 
for $W^{\pm}l^{\mp}$ decays democratic flavour dependence is assumed. The 
``Model 0'' \cite{gxx09} background is used, and for comparison with 
Fermi LAT data 25\% energy resolution is taken into account. From \cite{bix09}.
}
\label{fig:pf2}
\hskip -1cm
\end{figure}

Figure~\ref{fig:pf1} also shows the predicted total electron $+$ positron 
flux together with the results from Fermi LAT \cite{fermi09} and ATIC 
\cite{atic08}. Obviously, the ``Model 0''
cannot account for the present data, and the contribution
from gravitino decays makes the discrepancy even worse. In particular, the
data show no spectral feature expected for decaying dark matter. On the
other hand, gravitino decays may very well be consistent with the measured
total electron $+$ positron flux once the background is appropriately
adjusted. This is evident from Figure~\ref{fig:pf2} where the contribution
from gravitino decays is shown in the theoretically well motivated case
of flavour democratic decays. The figure also illustrates that, depending
on the gravitino mass, the dark matter contribution to the PAMELA excess
can still be significant.

An obvious possibility is that both, the total electron + positron flux and
the positron fraction, are dominated by astrophysical sources. For instance,
for the gravitino mass $m_{3/2} = 100$~GeV one obtains from the antiproton flux
constraint the minimal lifetime  $\tau_{3/2}^{\rm min}(100) \simeq 
1\times 10^{27}$~s. The corresponding contribution from gravitino decays
to the total electron + positron flux and positron fraction turns out to be 
indeed negligible \cite{bix09}. Nevertheless, as discussed in the next section,
the dark matter contribution to the gamma-ray flux can still be sizable.

\section{PREDICTIONS FOR THE GAMMA-RAY SPECTRUM}

The gamma-ray flux from gravitino dark matter decay receives contributions 
from the decay of gravitinos in the Milky Way halo and at cosmological 
distances, which can be calculated in the standard manner.
The halo component dominates, leading to a slightly anisotropic gamma-ray 
flux. 

The gravitino decay produces a continuous spectrum of gamma-rays
which is determined by the fragmentation of the Higgs boson and the weak gauge 
bosons. In addition, there exists a gamma-ray line at the endpoint of the 
spectrum with an intensity which is model-dependent. For our numerical analysis
we use a typical branching ratio in this channel,
\begin{equation}
\mathrm{RR}(\psi_{3/2}\rightarrow \nu\gamma) = 
0.02 \left(\frac{200~\mathrm{GeV}}{m_{3/2}}\right)^2 , \nonumber
\end{equation}
for gravitino masses above $100$~GeV.
In Figure~\ref{fig:gamma} the predicted diffuse gamma-ray
flux is shown for $m_{3/2} = 100,~200~\mathrm{GeV}$ and the
respective lower bounds on the gravitino lifetime. These 
spectra correspond to upper bounds on the 
signal in gamma-rays that can be expected from gravitino dark matter decay. 

For our analysis, two sets of results are used since 
the status of the extragalactic background is currently unclear. 
For the background obtained by Moskalenko et al., 
the extragalactic component is described by the power law \cite{smr04}
\begin{equation}
\left[E^2 \frac{dJ}{dE}\right]_\mathrm{bg}=6.8\times 10^{-7}
\left(\frac{E}{\rm GeV}\right)^{-0.32}
(\mathrm{cm}^2~\mathrm{str}~\mathrm{s})^{-1}~\mathrm{GeV} . 
\end{equation}
The earlier analysis by Sreekumar et al. led to
a less steep background \cite{sxx97},
\begin{equation}
\left[E^2 \frac{dJ}{dE}\right]_\mathrm{bg}=1.37\times 10^{-6}
\left(\frac{E}{\rm GeV}\right)^{-0.1}
(\mathrm{cm}^2~\mathrm{str}~\mathrm{s})^{-1}~\mathrm{GeV} . 
\end{equation}
In Figure~\ref{fig:gamma}, the slightly anisotropic halo signal has 
been averaged over the whole sky, excluding a band of $\pm 10^\circ$ around 
the Galactic disk. For the energy resultion $\sigma(E)/E=15\%$ has been used. 

\begin{figure}
\hskip -1.5cm
\includegraphics[height=.25\textheight]{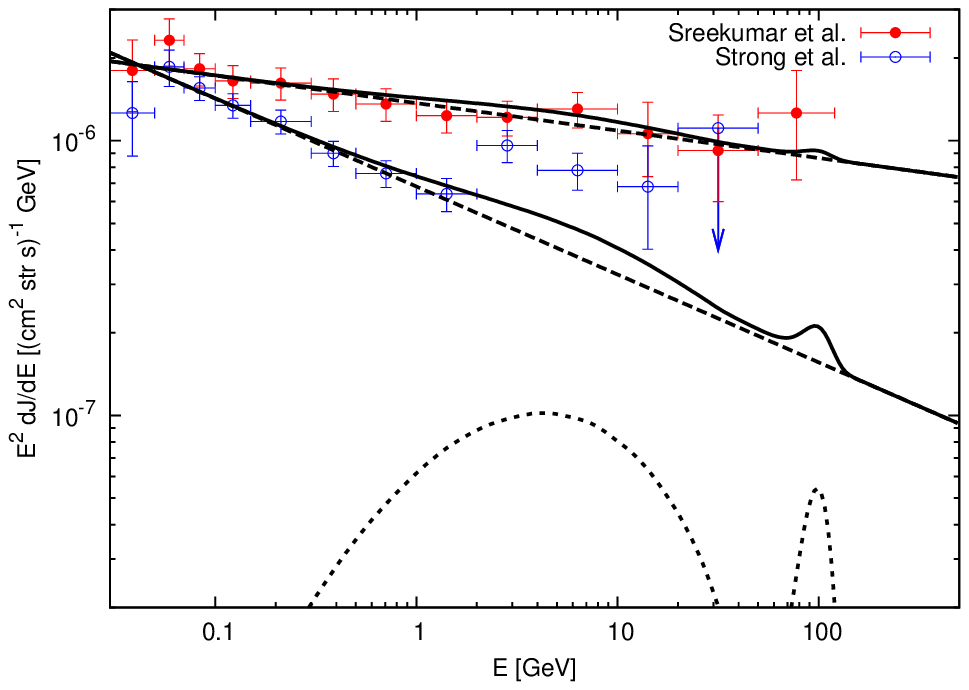}
\hskip 1cm
\includegraphics[height=.25\textheight]{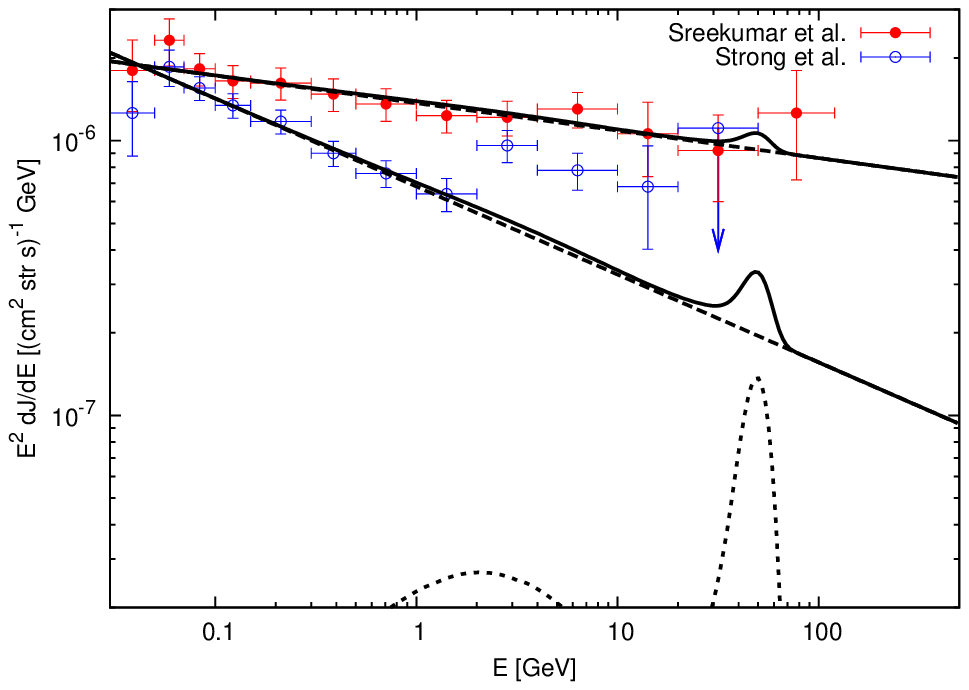}
\hskip -10.5cm {\small (a)}\hskip 8.5cm{\small (b)}
\caption{
Gamma-ray flux for (a) $m_{3/2} = 200~\mathrm{GeV}$,
$\tau_{3/2}^{\mathrm{min}} = 7\times 10^{26}$~s, and (b)
$m_{3/2} = 100~\mathrm{GeV}$, $\tau_{3/2}^{\mathrm{min}} = 1\times 10^{27}$~s. 
The signal is added to two different backgrounds obtained in 
\cite{smr04}, \cite{sxx97}. From \cite{bix09}.
}
\label{fig:gamma}
\end{figure}

\section{Conclusion}

In supersymmetric theories with small R-parity breaking thermally produced
gravitinos can account for the observed dark matter, consistent with 
leptogenesis and nucleosynthesis. Gravitino decays then contribute to 
antimatter cosmic rays as well as gamma-rays. Gravitino masses 
below 600~GeV are consistent with universal boundary conditions at the
GUT scale. 

Gravitino decays into Standard Model particles can be studied in a 
model-independent way by means of an operator analysis. For sufficiently
large gravitino masses the dimension-5 operator dominates. This means
that the branching ratios into $h\nu$, $Z\nu$ and $W^{\pm}l^{\mp}$ are
fixed. As a consequence,
the gamma-ray flux is essentially determined once the antiproton flux is known.
On the contrary, the positron flux is model-dependent.
The gamma-ray line is controlled by the dimension-6 operator. Its
strength is model-dependent and decreases with increasing gravitino mass. 

Electron and positron fluxes from gravitino decays cannot account for both, 
the PAMELA 
positron fraction and the electron $+$ positron flux measured by Fermi LAT. 
For gravitino dark matter, the observed fluxes require astrophysical sources.  
However, depending on the gravitino mass, the dark matter 
contribution to the electron and positron fluxes can be non-negligable.

Present data on antiproton cosmic-rays allow for a sizable contribution of
gravitino dark matter to the gamma-ray spectrum, in particular a line
at an energy below 300~GeV. Non-observation of such a line would place
a lower bound on the gravitino lifetime, and hence on the strength
of R-parity breaking, restricting possible signatures at the LHC.

\section{Acknowledgments}
This talk is based on recent work with A.~Ibarra, M.~Endo, T.~Shindou,
F.~Takayama and D.~Tran whom I thank for a fruitful collaboration.        

\newpage


\end{document}